\documentclass[%
aip,
 amsmath,amssymb,
 reprint,%
]{revtex4-1}

\usepackage{graphicx}
\usepackage{dcolumn}
\usepackage{bm}
\usepackage{ulem}
\usepackage{threeparttable}
\usepackage{adjustbox}
\usepackage{color}

\DeclareRobustCommand{\katoerase}{\bgroup\markoverwith{\textcolor[rgb]{0,0.6,0}{\rule[.5ex]{2pt}{0.8pt}}}\ULon}

\usepackage[utf8]{inputenc}
\usepackage[T1]{fontenc}
\usepackage{mathptmx}

\begin{document}

\preprint{manuscript for {\it Journal of Rheology}}

\title[]{Development of Rheological Constitutive Modeling Method Using \\a Sparse Identification Algorithm: A Case Study for Extensional Flows}

\author{Takeshi Sato}
\email{takeshis@se.kanazawa-u.ac.jp}
 \affiliation{
Advanced Manufacturing Technology Institute, Kanazawa University, Kanazawa 920-1192, Japan
 }

\author{Souta Miyamoto}%
\affiliation{Department of Chemical Engineering, Graduate School of Engineering, Kyoto University, Kyoto 615-8510, Japan}

\author{Shota Kato}
\affiliation{Department of Informatics, Kyoto University, Kyoto 606-8501, Japan}

\date{\today}

\begin{abstract}
Deriving constitutive models (CMs) from numerical data has been an attractive approach as a systematic CM building method. One recent study is {\it Rheo}-SINDy, which extended the sparse identification of nonlinear dynamics (SINDy) method to rheology. Although the {\it Rheo}-SINDy framework discovered an approximate CM from numerical data under shear flow, its versatility has not been investigated. To clarify its applicability to other types of flows, this study applied {\it Rheo}-SINDy to numerically generated data under extensional flow conditions. As baseline tests for extensional flow, we considered two problems: (i) whether the {\it Rheo}-SINDy framework can reproduce the famous Giesekus model from data generated by that model, and (ii) whether it can derive an approximate CM from data generated by a dumbbell model with a finite extensible nonlinear elastic (FENE) spring. For problem (i), we confirmed that {\it Rheo}-SINDy can identify the exact expression of the Giesekus model under extensional flow. For problem (ii), the {\it Rheo}-SINDy framework discovered a relatively simple expression of the approximate CM by manually designing the library matrix based on rheological knowledge. The identified approximate CM can reasonably predict extensional rheological properties of the FENE dumbbell model, including an extrapolation region. These findings demonstrate the fundamental validity of using {\it Rheo}-SINDy under extensional flow. \\
\textbf{Key Words:} \\Viscoelastic fluids / Mesoscopic model / Constitutive modeling / Data-driven method / Sparse identification
\end{abstract}
\maketitle
%
\section{\label{introduction}INTRODUCTION}
Extensional flow is one of the fundamental flow modes in rheology, along with shear flow. Thus, understanding the extensional rheology of complex fluids (e.g., polymer and surfactant solutions) is a central topic in the rheological community. Experimentally, the extensional rheological properties are measured with, for example, filament-type or microfluidic devices.~\cite{Bach2003,McKinley2002,Dinic2015,Haward2023a,Haward2023b} In the filament-type devices, the filament stretching extensional rheometer (FiSER) can maintain a constant extensional rate and provide fundamental rheological data.~\cite{Bach2003} Capillary breakup extensional rheometry (CaBER)~\cite{McKinley2002} and dripping-onto-substrate (DoS) rheometry~\cite{Dinic2015} are techniques for observing the thinning behavior of a liquid filament driven by surface tension. These protocols are relatively easy to set up and have been used in many previous studies. On the other hand, measurements using microfluidic devices are more advanced, and there are relatively limited examples of research on this technology. For instance, Haward and coworkers have conducted extensional measurements using stagnation points formed within the channel with a highly optimized channel geometry, known as the optimized uniaxial and biaxial extensional rheometer (OUBER).~\cite{Haward2023a,Haward2023b} The OUBER device enables the measurement of steady extensional properties of dilute polymer solutions. These experimental measurement techniques continue to improve, contributing to a better understanding of extensional rheology and providing a baseline for modeling. 

Based on these measurements, researchers have attempted to predict extensional rheology using phenomenological constitutive or mesoscopic molecular models. The former models include, for example, the Giesekus~\cite{Giesekus1982} and Phan-Thien/Tanner (PTT)~\cite{Phan-Thien1977} models, which are straightforward to implement but generally lack (microscopic or mesoscopic) structural details. The latter models for polymers are based on coarse-grained molecular representations, such as the Rouse model for unentangled polymers~\cite{Rouse1953} and tube-based models for entangled polymers.~\cite{Doi1986} Molecular-based representations explicitly describe interactions between coarse-grained segments of polymers, thereby linking (coarse-grained) molecular dynamics with macroscopic rheological properties. These basic models led to several coarse-grained molecular models designed for numerical simulations of polymeric systems.~\cite{Masubuchi2001,Hua1998,Doi2003} Generally, these molecular models were first validated for their ability to represent equilibrium properties and then extended to non-equilibrium dynamics, including extensional rheology. Although molecular models have been extensively studied, a unified description of shear and extensional rheology remains a challenge, as summarized in several excellent review articles.~\cite{Ianniruberto2020,Matsumiya2021} The advanced theory of molecular rheology is expected to refine mesoscopic coarse-grained models, especially under strong flow, and research in this direction is currently underway.~\cite{Watanabe2021}

As a development in a different direction, data-driven science has been gradually adopted in rheology over the past few years.~\cite{Miyamoto2024} This trend provides one answer to the question ``How can we utilize data obtained from advanced experimental methods and sophisticated simulations?'' One of the pioneering studies of data-driven rheology is the rheology-informed neural networks (RhINNs) developed by Jamali and coworkers.~\cite{Mahmoudabadbozchelou2021} They employed a rheological constitutive equation to evaluate a loss function during training, thereby enabling the incorporation of rheological knowledge into the training model. Lennon and coworkers proposed a framework to develop the so-called rheological universal differential equations (RUDEs).~\cite{Lennon2023} RUDEs are designed using a tensor-based neural network (TBNN) and incorporate several physical principles, including frame invariance and tensor symmetry. Building on RUDEs, Sunol and coworkers recently proposed a strategy to derive constitutive equations from complex flow data.~\cite{Sunol2025} Molina and coworkers used a Gaussian process regression method, a Bayesian learning strategy, to develop a constitutive relation from data artificially generated by mesoscopic models.~\cite{Seryo2020,Seryo2021,Miyamoto2023} They successfully reproduced macroscopically inhomogeneous flows using their machine-learned constitutive relations. While the above approaches employ relatively complex training models (with a large number of parameters), we have recently proposed a method for deriving data-driven constitutive models (CMs) with a (simple) sparse identification algorithm.~\cite{Sato2025a} (It is fair to note that Shanbhag and Erlebacher also used the sparse identification algorithm to derive CMs.~\cite{Shanbhag2024}) We extended the so-called sparse identification of nonlinear dynamics (SINDy)~\cite{Brunton2016} to derive CMs that can predict shear-dominated flow properties within and (moderately) outside the training data.~\cite{Sato2025b}

This study examined the ability of our data-driven strategy, rheological constitutive modeling with SINDy ({\it Rheo}-SINDy), to model extensional rheology. For this purpose, we test (i) whether {\it Rheo}-SINDy can identify a correct model from data generated from a phenomenological CM (i.e., the Giesekus model~\cite{Giesekus1982}) and (ii) whether it can derive an approximate CM from a basic mesoscopic model. In (ii), we employ a dumbbell model with a finitely extensible nonlinear elastic (FENE) spring, which lacks an analytical CM.~\cite{Bird1987} Although the dumbbell model is the basic model for viscoelastic fluids, it is well-suited for testing advanced theories and incorporating previously unaccounted physical effects.~\cite{Watanabe2020,Uneyama2021} We attempt to derive an approximate CM for the FENE dumbbell model, incorporating our rheological knowledge. Details are shown below. 
\section{\label{data-driven-method}DATA-DRIVEN METHOD}
This section provides a brief explanation of our strategy to obtain (approximate) CMs, which extended one of the symbolic regression algorithms (i.e., the SINDy algorithm).~\cite{Sato2025a,Sato2025b} The SINDy algorithm provides a symbolic representation of a governing differential equation from dynamical data. 

\begin{figure}[t]
\centering
\includegraphics[width=0.9\columnwidth]{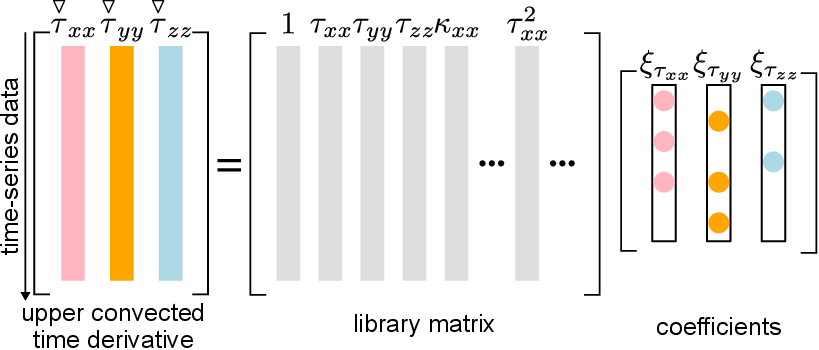}
\caption{Schematic representation of the {\it Rheo}-SINDy framework.}
\label{Fig01}
\end{figure}

Figure~\ref{Fig01} shows the framework of our data-driven strategy. Assuming that the upper convected derivative expresses the derivative terms in CMs, we can write the general expression for CMs as $\overset{\triangledown}{\boldsymbol \tau} = {\boldsymbol f} ({\boldsymbol \tau},{\boldsymbol \kappa})$, where $\overset{\triangledown}{\boldsymbol \tau}$ $(= {\rm d} {\boldsymbol \tau} (t) / {\rm d}t - {\boldsymbol \tau} \cdot {\boldsymbol \kappa}^{\rm T} - {\boldsymbol \kappa} \cdot {\boldsymbol \tau})$ is the upper convected derivative of the stress tensor $\boldsymbol \tau$, $\boldsymbol \kappa$ is the velocity gradient tensor, and ${\boldsymbol \kappa}^{\rm T}$ is the transpose of ${\boldsymbol \kappa}$. Using the general expression for CMs, we obtain the expression of CMs for the SINDy algorithm as
\begin{equation}
\overset{\triangledown} {\boldsymbol T} = {\boldsymbol \Theta} \left ( {\boldsymbol T}, {\boldsymbol K} \right ) \boldsymbol \Xi,
\label{SINDy_Matrix}
\end{equation}
where ${\boldsymbol T}$ and ${\boldsymbol K}$ are the time-series training sets for the stress and velocity gradient, respectively, ${\boldsymbol \Theta} \left ( {\boldsymbol T}, {\boldsymbol K} \right )$ is the library matrix whose components are functions of ${\boldsymbol T}$ and ${\boldsymbol K}$, and $\boldsymbol \Xi$ is the coefficient matrix. Since this study examines the extensional flow, the column elements of the matrices ${\boldsymbol T}$ and ${\boldsymbol K}$ can be summarized as ${\boldsymbol T} = \left [ {\boldsymbol t}_{xx}~{\boldsymbol t}_{yy}~{\boldsymbol t}_{zz}  \right ]$ and ${\boldsymbol K} = \left [ {\boldsymbol k}_{xx}~{\boldsymbol k}_{yy}~{\boldsymbol k}_{zz}  \right ]$, where ${\boldsymbol t}_{ii}$ $(=\left [ \tau_{ii} (t_1)~\tau_{ii} (t_2)~\cdots~\tau_{ii} (t_n)\right ]^{\rm T})$ and ${\boldsymbol k}_{ii}$ $(=\left [ \kappa_{ii} (t_1)~\kappa_{ii} (t_2)~\cdots~\kappa_{ii} (t_n)\right ]^{\rm T})$ are the stress and velocity gradient data, respectively, with the subscript $i$ denoting $\{x,y,z\}$. 
${\boldsymbol \Theta} \left ( {\boldsymbol T}, {\boldsymbol K} \right )$ can include any functions made by ${\boldsymbol T}$ and ${\boldsymbol K}$ and can incorporate prior rheological knowledge. 

We determine non-zero components of $\boldsymbol \Xi$ using a sparsity-promoting regression algorithm. In general, the solution depends on the regression algorithm employed.
Such regression problems are typically formulated as optimization problems with regularization terms. The basic choice of the regularization term is the $\ell_1$ or $\ell_2$ norm. The method with the former is called the least absolute shrinkage and selection operator (Lasso) regression, whereas the method with the latter is called the Ridge regression. Our previous study revealed that the more advanced methods based on these basic regression methods, namely adaptive Lasso (a-Lasso)~\cite{Zou2006,Fukami2021} and sequentially thresholded Ridge (STRidge)~\cite{Rudy2017} regressions, are effective for our purpose.~\cite{Sato2025a} Note that STRidge promotes sparsity by the thresholding algorithm, not by the $\ell_2$ norm. Thus, this study tested these two regression methods. Each of these methods includes a hyperparameter, $\alpha$, that controls the sparsity of the solution. We used the following normalized total mean squared error (MSE) to determine a solution from a wide range of $\alpha$ values: 
\begin{equation}
{\overline {\rm MSE}} = w {\overline {\rm MSE}}_{\rm training} + (1-w) {\overline {\rm MSE}}_{\rm solved}, 
\label{MSE}
\end{equation}
where $w$ is the weight factor ($w = 0.5$), ${\overline {\rm MSE}}_{\rm training}$ is the MSE between $\overset{\triangledown} {\boldsymbol T}$ and the reconstructed data, normalized by the value when the number of terms is $0$ (i.e., the sparsest case), and ${\overline {\rm MSE}}_{\rm solved}$ is the MSE between ${\boldsymbol T}$ and the stress data obtained by integrating the identified CMs, also normalized by the value when the number of terms is $0$. Our original study used only ${\overline {\rm MSE}}_{\rm training}$ ($w = 1$).~\cite{Sato2025a} We confirmed that when $w = 1$, a solution with a smaller $\alpha$ (i.e., a solution with a larger number of terms) that better fits the training data ($\overset{\triangledown} {\boldsymbol T}$) tends to have the smallest MSE, and that this solution is selected as optimal when we rely solely on ${\overline {\rm MSE}}_{\rm training}$ to obtain a final model. Furthermore, the selected model that relies solely on ${\overline {\rm MSE}}_{\rm training}$ may fail to yield bounded solutions; for example, the model containing terms that lead to unphysical responses will diverge during numerical integration. ${\overline {\rm MSE}}_{\rm solved}$ helps exclude such models and identify integrable models. Our preliminary tests revealed that setting $w < 1$ gives a numerically stable solution, consequently yielding a sparse model that provides reasonable stress predictions. The qualitative results remained unchanged when $w \le 0.5$; thus, we set $w = 0.5$ for the following tests. The (nearly) optimal solution is determined by using the MSE in Eq.~\eqref{MSE}. Since the sparse identification strategy favors a parsimonious solution, if the magnitude of the MSE is roughly at the same level over a specific range of $\alpha$, the sparser solution is considered optimal.
\section{RHEOLOGICAL MODELS}
This section introduces the rheological models used to generate the data in this study, namely the Giesekus and FENE dumbbell models. For simplicity, we used a single-mode representation for both models; thus, the linear rheological properties are characterized by a pair of relaxation time $\lambda$ and modulus $G$. 

We used uniaxial, planar, and biaxial extensional flows to generate the training data. Each velocity gradient tensor ($\boldsymbol \kappa$) is expressed as 
\begin{equation}
\boldsymbol \kappa_{\rm UE} = 
\begin{pmatrix} 
  \dot \epsilon & 0 & 0 \\
  0 & -\dot \epsilon/2 & 0 \\
  0 & 0 & -\dot \epsilon/2
\end{pmatrix},
\label{UE}
\end{equation}
\begin{equation}
\boldsymbol \kappa_{\rm PE} = 
\begin{pmatrix} 
  \dot \epsilon & 0 & 0 \\
  0 & -\dot \epsilon & 0 \\
  0 & 0 & 0
\end{pmatrix},
\label{PE}
\end{equation}
and 
\begin{equation}
\boldsymbol \kappa_{\rm BE} = 
\begin{pmatrix} 
  \dot \epsilon & 0 & 0 \\
  0 & \dot \epsilon & 0 \\
  0 & 0 & -2 \dot \epsilon
\end{pmatrix}, 
\label{BE}
\end{equation}
where $\dot \epsilon$ is the extensional rate. As shown in Eqs.~\eqref{UE}--\eqref{BE}, the velocity gradient tensor is traceless due to the incompressibility. We note that the flow field corresponds to uniaxial extensional flow in some phases and to biaxial extensional flow in other phases when the extension rate $\dot \epsilon (t)$ oscillates. Carefully designing training datasets that account for flow-field characteristics is one of our important future works.
\subsection{Giesekus model}
In addition to the terms that appear in the most basic upper-convected Maxwell (UCM) model for viscoelastic fluids, the Giesekus model includes a quadratic stress term to reproduce the nonlinear rheological properties. 
Using the unit time $\lambda$ and stress $G$, we can obtain the dimensionless expression of the Giesekus model as
\begin{equation}
\overset{\triangledown}{\boldsymbol \tau} = - \boldsymbol \tau - \alpha_{\rm G} \boldsymbol \tau \cdot \boldsymbol \tau + 2 \boldsymbol D,
\label{Giesekus}
\end{equation}
where $\alpha_{\rm G}$ is the parameter to control the nonlinear rheological properties, and $\boldsymbol D$ is the strain rate tensor defined as $\boldsymbol D \equiv (\boldsymbol \kappa + {\boldsymbol \kappa}^{\rm T})/2$. Under extensional flow, where only the diagonal components of $\boldsymbol \tau$ and $\boldsymbol D$ are important, Eq.~\eqref{Giesekus} reduces to 
\begin{equation}
\overset{\triangledown}{\tau}_{ii} = - \tau_{ii} - \alpha_{\rm G} \tau^2_{ii} + 2 \kappa_{ii},  
\label{Giesekus_Component}
\end{equation}
where the subscript $i$ denotes the coordinate direction (i.e., $xx$, $yy$, and $zz$). 

We generated the training data by numerically integrating Eq.~\eqref{Giesekus_Component} ($\alpha_{\rm G} = 0.3$) under oscillatory uniaxial and biaxial extensional flows with different strain amplitudes. As can be inferred from Eqs.~\eqref{UE}--\eqref{BE}, we cannot correctly identify the equations for the three independent stresses if only one extensional mode is used for training. For example, under uniaxial extensional flow, since $\kappa_{yy} = \kappa_{zz}$, we cannot distinguish between the $yy$ and $zz$ components of the stress responses. Additionally, we found that the combined data of uniaxial ($\kappa_{yy} = \kappa_{zz}$) and planar ($\kappa_{zz} = 0$) extensional flows did not contain enough information to identify the correct expression of the equation for $\overset{\triangledown}{\tau}_{zz}$ in the current setting. Based on these observations, we decided to use the two extensional modes: uniaxial and biaxial extensional flows.
The time-dependent extensional rate $\dot \epsilon (t)$, non-dimensionalized by $\lambda$, is determined as $\dot \epsilon (t) = \epsilon_0 \omega \cos (\omega t)$, where $\epsilon_0$ is the strain amplitude and $\omega$ is the angular frequency. Following the training strategy developed by Miyamoto and coworkers,~\cite{Miyamoto2023} we recorded the diagonal components of $\boldsymbol \tau$ up to $t = 10$ at a time interval of $\Delta t = 0.01$, varying $\epsilon_0 = 2$, $4$, and $6$ for the uniaxial extensional flow and $\epsilon_0 = 1$, $2$, and $3$ for the biaxial extensional flow while keeping $\omega = 1$. Thus, each extensional mode has $N_{\rm train} = 3 \times 10^3$ time-series samples for training. We set these $\epsilon_0$ values to align the stress magnitude levels of uniaxial and biaxial extensions. We included in the library matrix $\boldsymbol \Theta (\boldsymbol T, \boldsymbol K)$ polynomials up to the third degree that can be constructed from the six descriptors $\{ \tau_{xx},\tau_{yy},\tau_{zz},\kappa_{xx},\kappa_{yy},\kappa_{zz} \}$. Thus, the number of candidates to identify a time-evolution equation for each stress component is $N_{\boldsymbol \Theta} = 84$. 
\subsection{FENE dumbbell model}
The FENE dumbbell model is the fundamental mesoscopic model for viscoelastic fluids. In this model, a polymer chain is coarse-grained into two beads, denoted by ${\boldsymbol r}_1 (t)$ and ${\boldsymbol r}_2 (t)$, connected by a spring. The following Langevin equation expresses the time evolution of the end-to-end vector connecting the two beads $\boldsymbol R (t)$: 
\begin{equation}
\zeta \left \{ \frac{{\rm d} {\boldsymbol R} (t)}{{\rm d} t} - \boldsymbol \kappa \cdot {\boldsymbol R} (t) \right \} = - 2 h(t) {\boldsymbol R} (t) + {\boldsymbol F}_{21}^{\rm B}(t),
\label{Dumbbell}
\end{equation}
where $\zeta$ is the friction coefficient, $h(t)$ is the spring strength, and ${\boldsymbol F}_{21}^{\rm B}(t)$ $(= {\boldsymbol F}_{2}^{\rm B}(t) - {\boldsymbol F}_{1}^{\rm B}(t))$ is the difference in the thermal fluctuation force acting on beads 1 and 2, denoted by ${\boldsymbol F}_{1}^{\rm B}(t)$ and ${\boldsymbol F}_{2}^{\rm B}(t)$, respectively. Equation~\eqref{Dumbbell} indicates that our dumbbell model accounts for frictional, elastic, and Brownian forces. The spring strength is expressed as
\begin{equation}
h (t) = h_{\rm eq} f_{\rm FENE} (t),
\label{spring_strength}
\end{equation}
where $h_{\rm eq}$ is the spring strength at equilibrium and $f_{\rm FENE} (t)$ is the FENE factor defined as 
\begin{equation}
f_{\rm FENE} (t) = \frac{1}{1 - {\boldsymbol R}^2(t)/R_{\rm max}^2}, 
\label{FENE}
\end{equation}
with $R_{\rm max}$ being the maximum length of the bead's end-to-end vector. Using $\boldsymbol R(t)$ evaluated by Eqs.~\eqref{Dumbbell}--\eqref{FENE}, we can compute ${\boldsymbol \tau} (t)$ as
\begin{equation}
{\boldsymbol \tau} (t) = \frac{3G}{R_{\rm eq}^2} \langle f_{\rm FENE}(t) {\boldsymbol R}(t){\boldsymbol R}(t) \rangle - G {\boldsymbol I}. 
\label{dumbbell_stress}
\end{equation}
Here, $R_{\rm eq}$ is the length of the bead's end-to-end vector at equilibrium. The above equations for the FENE dumbbell mode have been made dimensionless by taking the unit time as the relaxation time $\lambda = \zeta / (4h_{\rm eq})$, the unit length as the equilibrium dumbbell length $R_{\rm eq}$, and the unit stress as the modulus $G$.

Since the FENE dumbbell model cannot yield an analytical CM, we relied on Brownian dynamics (BD) simulations with a small time step size $(\leq 10^{-4})$ to generate training data under oscillatory uniaxial, planar, and biaxial extensional flows with different $\epsilon_0$ values while fixing $\omega = 1$. Through this study, we set $R_{\rm max} = 3 R_{\rm eq}$. As in the Giesekus model, we applied the time-dependent $\dot \epsilon (t)$ with $\epsilon_0 \in \{2,4,6,8\}$ for the uniaxial and planar extensional flows and $\epsilon_0 \in \{1,2,3,4\}$ for the biaxial extensional flow. Since we recorded the diagonal components of $\boldsymbol \tau$ up to $t = 10$ at a time interval of $\Delta t = 0.01$, each extensional mode has $N_{\rm train} = 4 \times 10^3$ time-series samples for training. 

This study explores an approximate CM of the FENE dumbbell model under extensional flow using the {\it Rheo}-SINDy framework (cf. Eq.~\eqref{SINDy_Matrix}). To provide a practical expression for the problem without an analytical solution, we adopt a strategy of designing the library matrix $\boldsymbol \Theta (\boldsymbol T, \boldsymbol K)$ using our rheological knowledge.
For this purpose, we used the FENE-P dumbbell model, which is analytically more tractable than the FENE dumbbell model. The FENE-P dumbbell model enables analytical treatment by using a pre-averaging approximation in Eq.~\eqref{FENE} (cf. Eq.~\eqref{FENE_P} in Appendix~\ref{App_A}). Although the FENE-P dumbbell model can be simply expressed using the conformation tensor (cf. Eqs.~\eqref{fp_dumbbell_stress} and \eqref{C} in Appendix~\ref{App_A}), we adopt a stress-based CM expression, obtained using the method of Mochimaru~\cite{Mochimaru1983} and summarized in the Appendix~\ref{App_A}. We utilized all terms in Eq.~\eqref{FENE_P_stress_expression_final} to design $\boldsymbol \Theta (\boldsymbol T, \boldsymbol K)$; thus, the total number of candidate terms per equation for the stress component is $N_{\boldsymbol \Theta} = 16$. 
\begin{figure}[!htbp]
\centering
\includegraphics[width=0.9\columnwidth]{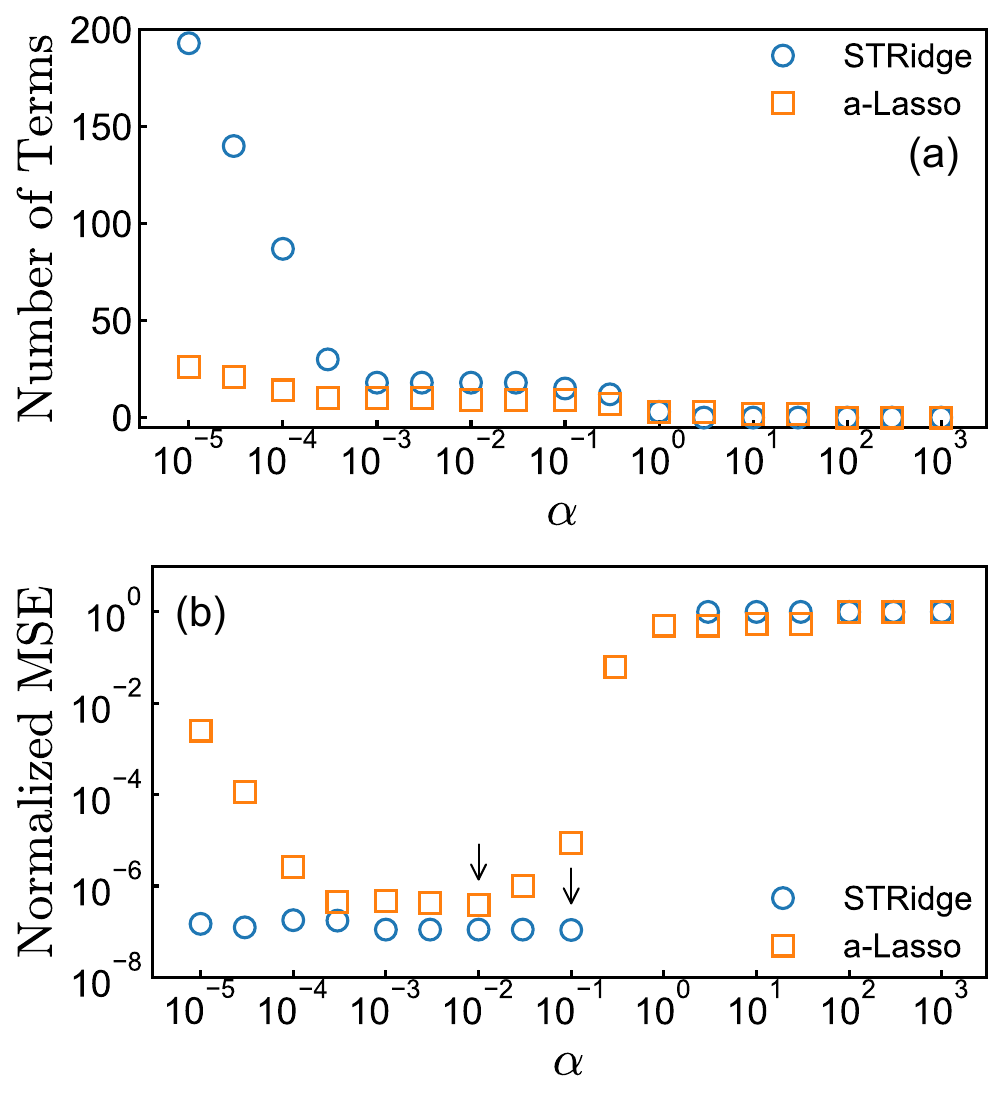}
\caption{Hyperparameter ($\alpha$) dependence of (a) the total number of identified terms of the equations for $\overset{\triangledown}{\tau}_{xx}$, $\overset{\triangledown}{\tau}_{yy}$, and $\overset{\triangledown}{\tau}_{zz}$, and (b) the normalized MSE, obtained for the Giesekus model. We obtained these results from the training data, which include the data generated under uniaxial and biaxial extensional flows. The circles and squares correspond to the results obtained by STRidge and a-Lasso, respectively. In (b), values of $\alpha$ with no plotted symbols indicate the cases where the identified CM diverged or where the MSE was larger than the sparsest case ($\overset{\triangledown}{\boldsymbol \tau} = {\boldsymbol 0}$). The downward arrows in (b) indicate the (nearly) optimal $\alpha$ for each regression method.}
\label{Fig02}
\end{figure}
\begin{figure}[!htbp]
\centering
\includegraphics[width=0.9\columnwidth]{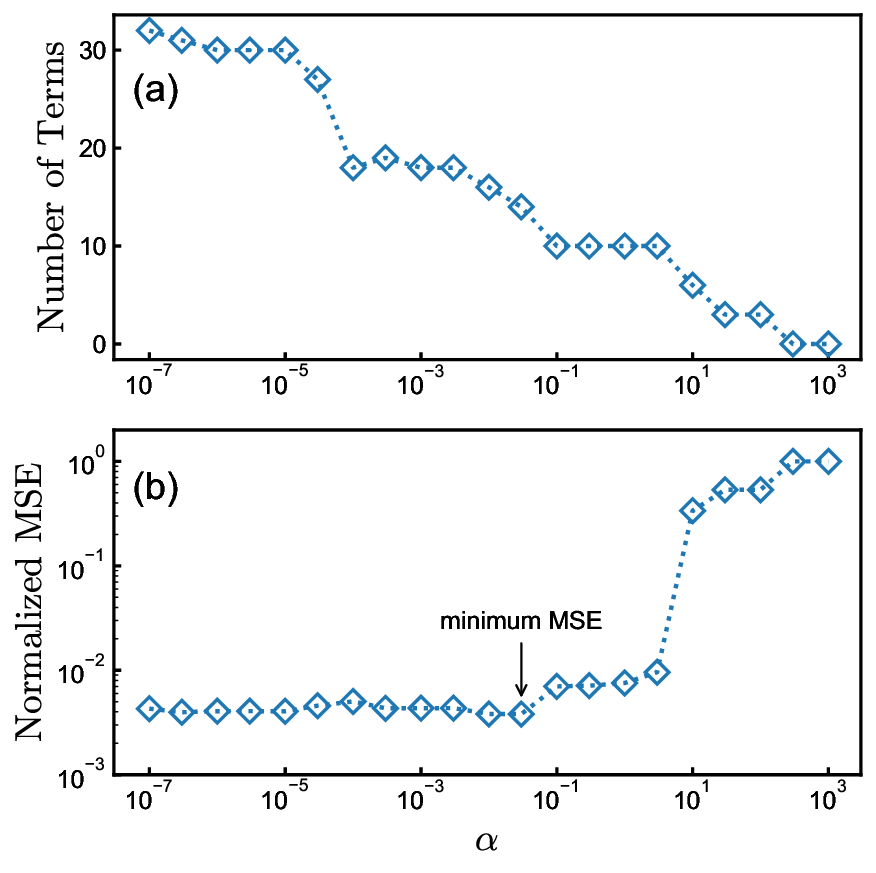}
\caption{Hyperparameter ($\alpha$) dependencies of (a) the total number of identified terms and (b) the normalized MSE, both obtained by a-Lasso for the FENE dumbbell model. The downward arrow in (b) indicates the (nearly) optimal $\alpha$ value that has a minimum MSE ($\alpha = 3 \times 10^{-2}$). The number of terms at $\alpha = 3 \times 10^{-2}$ is $14$. }
\label{Fig03}
\end{figure}
\begin{figure*}[t]
\centering
\includegraphics[width=0.9\textwidth]{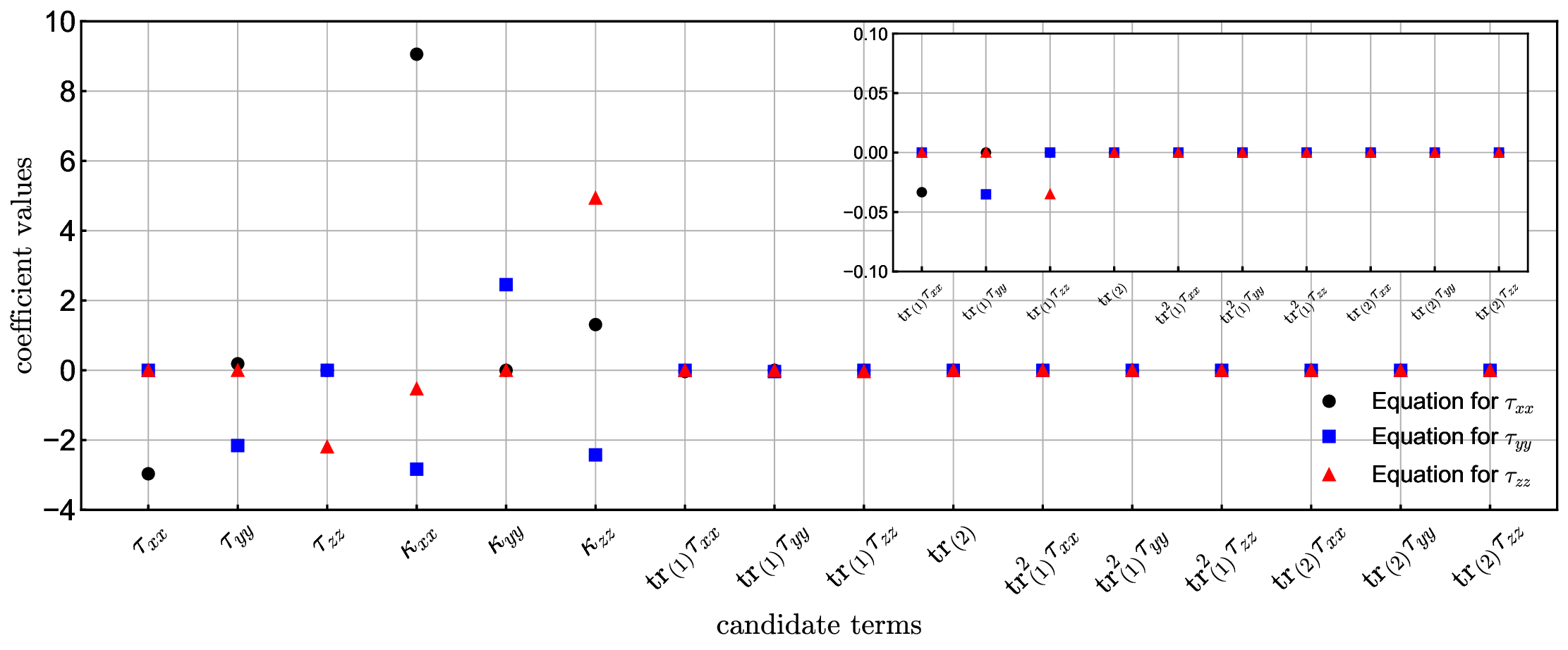}
\caption{The identified coefficient values for the FENE dumbbell model obtained by a-Lasso with $\alpha = 3 \times 10^{-2}$, which is the (nearly) optimum $\alpha$ value shown by the black arrow in Fig.~\ref{Fig03}(b). The black circles, blue squares, and red triangles show the coefficient values of the equations for ${\overset{\triangledown}\tau}_{xx}$, ${\overset{\triangledown}\tau}_{yy}$, and ${\overset{\triangledown}\tau}_{zz}$, respectively. The inset shows the enlargement of the coefficient values for higher-order terms. ${\rm tr}_{(1)}$ and ${\rm tr}_{(2)}$ indicate ${\rm tr}_{(1)} = {\rm tr} {\boldsymbol \tau}$ and ${\rm tr}_{(2)} = {\rm tr} ({\boldsymbol \tau} \cdot {\boldsymbol \kappa}^{\rm T} + {\boldsymbol \kappa} \cdot {\boldsymbol \tau})$, respectively.}
\label{Fig04}
\end{figure*}
\begin{figure}[b]
\centering
\includegraphics[width=0.9\columnwidth]{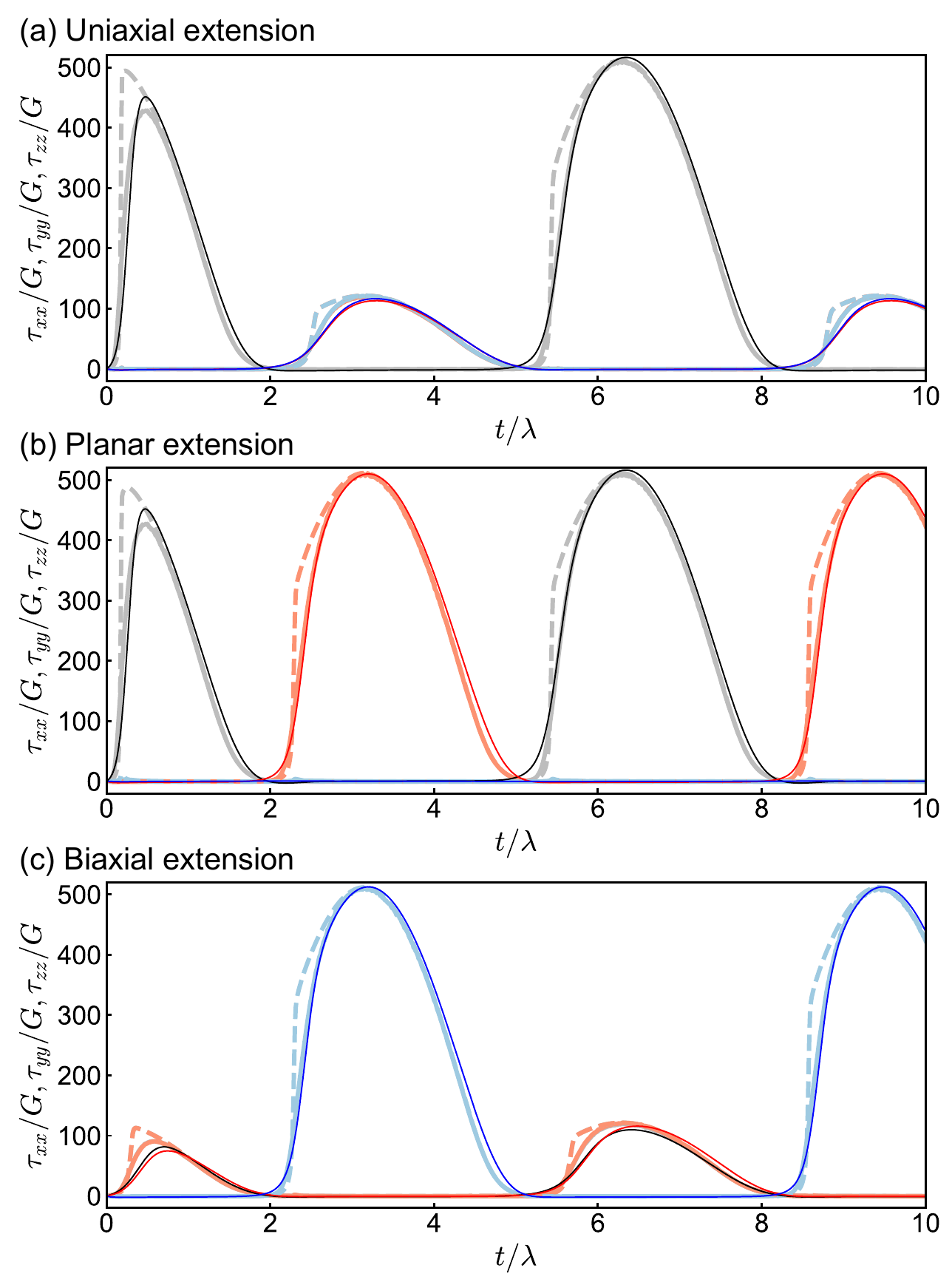}
\caption{The test simulation results for the oscillatory (a) uniaxial, (b) planar, and (c) biaxial extensional flows with $\omega = 1$. Here, the strain amplitude was $\epsilon_0 = 10$ for the uniaxial and planar extensional flows, and $\epsilon_0 = 5$ for the biaxial extensional flow. The thin solid, bold solid, and bold dotted lines represent the results obtained using the identified CMs, the FENE dumbbell model, and the FENE-P dumbbell model, respectively. The black, red, and blue lines show $\tau_{xx}$, $\tau_{yy}$, and $\tau_{zz}$, respectively.}
\label{Fig05}
\end{figure}
\begin{figure}[b]
\centering
\includegraphics[width=0.9\columnwidth]{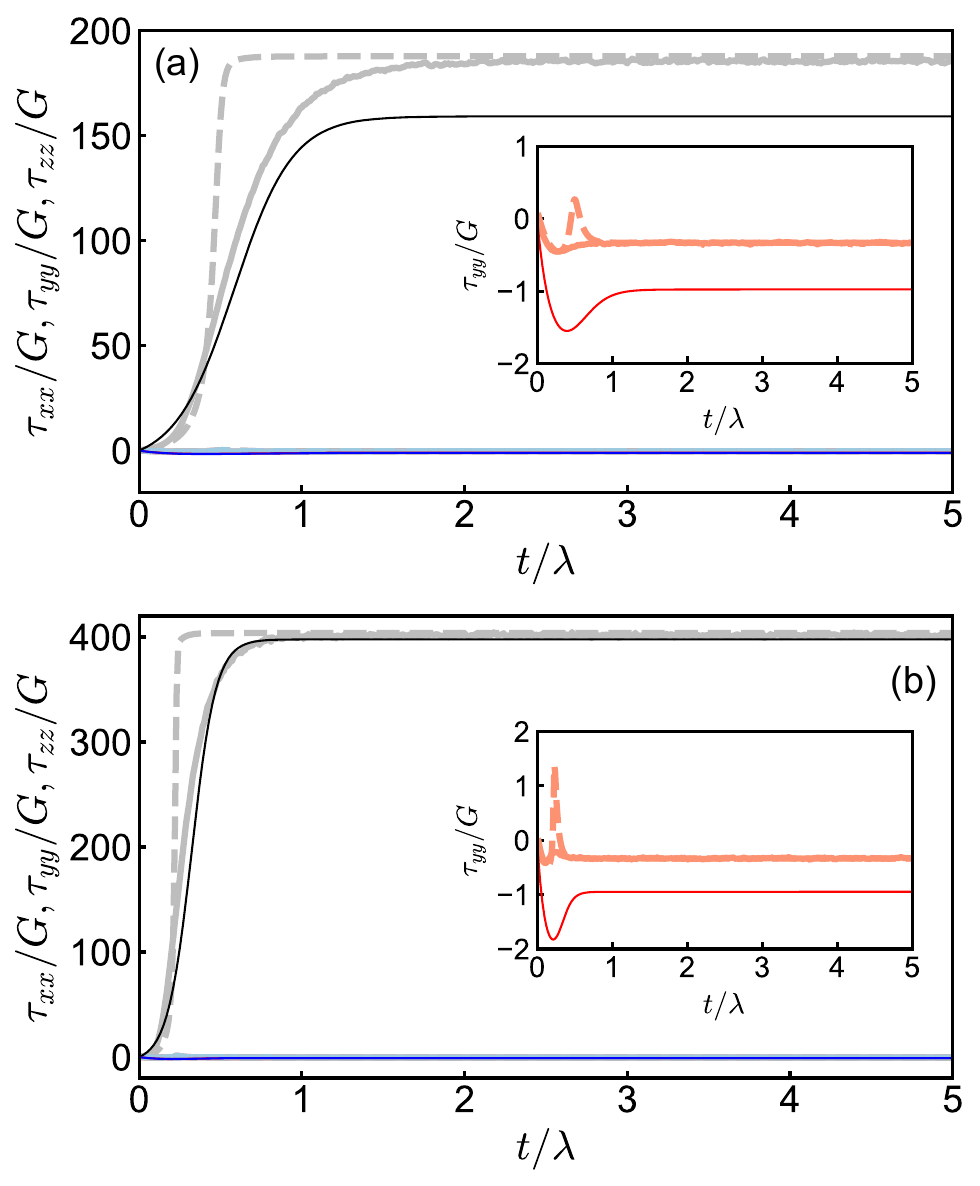}
\caption{The test simulation results for the uniaxial extensional flows with the steady extensional rate of (a) $\dot \epsilon = 4$ and (b) $8$. The thin solid, bold solid, and bold dotted lines represent the results obtained using the identified CMs, the FENE dumbbell model, and the FENE-P dumbbell model, respectively. The black, red, and blue lines show $\tau_{xx}$, $\tau_{yy}$, and $\tau_{zz}$, respectively. The inset shows an enlarged view of $\tau_{yy}$.}
\label{Fig06}
\end{figure}
\begin{figure}[t]
\centering
\includegraphics[width=0.9\columnwidth]{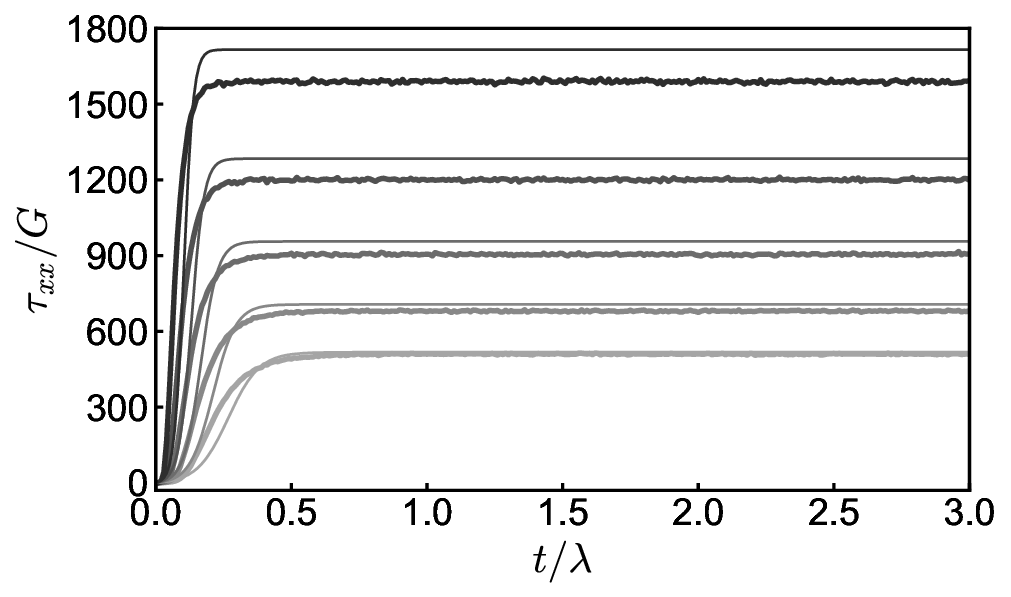}
\caption{The test simulation results for the uniaxial extensional flows at steady extensional rates of $\dot{\epsilon} = 10$, $13$, $17$, $23$, and $30$ (from bottom to top). Note that the extension rates are logarithmically spaced from $\dot \epsilon = 10$ to $\dot \epsilon = 30$.}
\label{Fig07}
\end{figure}
\section{\label{results_and_discussions}RESULTS AND DISCUSSION}
\subsection{Giesekus model}
Figure~\ref{Fig02} indicates the $\alpha$ dependence of the equation-identification results obtained from training data generated by uniaxial and biaxial extensional flows. Generally, a larger $\alpha$ yields a sparser solution and an increased MSE. The non-monotonicity of the MSE observed in the a-Lasso case shown in Fig.~\ref{Fig02}(b) suggests that the correct terms of the Giesekus model are identified in the range of $\alpha$ that shows a small MSE. Consistent with our previous works,~\cite{Sato2025a,Sato2025b} a-Lasso yields sparser results in a broad range of $\alpha$ values than STRidge. Note that this tendency in the difference between a-Lasso and STRidge depends on the parameter sets used in the a-Lasso and STRidge algorithms. Figure~\ref{Fig02}(b) determined the (nearly) optimal value of $\alpha$ for each regression method, as indicated by the black arrows. Here, we selected a (nearly) optimal $\alpha$ according to our criteria explained in Sec.~\ref{data-driven-method}. Here, we did not use the exact number of terms in the Giesekus model (cf. Eq.~\eqref{Giesekus_Component}) to determine the optimal $\alpha$. 

The identified equations (without any post-training correction) obtained by STRidge and a-Lasso are shown as 
\begin{align}
\overset{\triangledown}{{\tau}}_{xx} &= -1.001\tau_{xx} - 0.300\tau_{xx}^2 \nonumber \\
&\quad + 1.333 \kappa_{xx} - 0.666 \kappa_{yy} - 0.667 \kappa_{zz}, \label{G_str_xx}\\
\overset{\triangledown}{{\tau}}_{yy} &= -1.001\tau_{yy} - 0.300\tau_{yy}^2 \nonumber \\
&\quad - 0.667 \kappa_{xx} + 1.334 \kappa_{yy} - 0.667 \kappa_{zz}, \label{G_str_yy}\\
\overset{\triangledown}{{\tau}}_{zz} &= -1.001\tau_{zz} - 0.300\tau_{zz}^2 \nonumber \\
&\quad - 0.667 \kappa_{xx} - 0.667 \kappa_{yy} + 1.334 \kappa_{zz}, \label{G_str_zz}
\end{align}
and 
\begin{align}
\overset{\triangledown}{{\tau}}_{xx} &= -1.006\tau_{xx} - 0.300\tau_{xx}^2 + 2.001 \kappa_{xx}, \label{G_als_xx}\\
\overset{\triangledown}{{\tau}}_{yy} &= -1.009\tau_{yy} - 0.299\tau_{yy}^2 + 2.002 \kappa_{yy}, \label{G_als_yy}\\
\overset{\triangledown}{{\tau}}_{zz} &= -1.008\tau_{zz} - 0.300\tau_{zz}^2 + 2.003 \kappa_{zz}, \label{G_als_zz}
\end{align}
respectively. Equations~\eqref{G_als_xx}--\eqref{G_als_zz} show that a-Lasso (almost) accurately identifies the exact equation shown in Eq.~\eqref{Giesekus_Component}. Equations~\eqref{G_str_xx}--\eqref{G_str_zz} show that STRidge appears to fail to identify the correct equations. However, if we look carefully at the equations, the incompressible condition ${\rm tr} {\boldsymbol D} = {\rm tr} {\boldsymbol \kappa} = 0$ implies that $1.333 \kappa_{xx} - 0.666 \kappa_{yy} - 0.667 \kappa_{zz} \simeq  2 \kappa_{xx} - 2/3 \times (\kappa_{xx} + \kappa_{yy} + \kappa_{zz}) = 2 \kappa_{xx}$ in Eq.~\eqref{G_str_xx}, for example, which indicates STRidge has also identified the correct equations. Thus, both regression methods successfully yielded the Giesekus model. In general, since a-Lasso yields sparser solutions with roughly the same MSE as STRidge, we present only a-Lasso results for the more challenging FENE dumbbell model shown in the subsequent section.
\subsection{FENE dumbbell model}
Based on the success with the Giesekus model and the partial success with the FENE-P dumbbell model shown in Appendix~\ref{App_B}, we attempted to obtain an approximate CM for the FENE dumbbell model, which has no analytical CM. Figure~\ref{Fig03} presents the hyperparameter ($\alpha$) dependence of (a) the total number of terms and (b) the normalized MSE values, both obtained by a-Lasso. As shown in Fig.~\ref{Fig02} and in our previous studies,\cite{Sato2025a,Sato2025b} the number of terms decreases with increasing $\alpha$, whereas the MSE increases. From the examined range of $\alpha$ values, we chose the (nearly) optimal $\alpha$ that yields the minimum MSE ($\alpha = 3 \times 10^{-2}$). As shown in Fig.~\ref{Fig03}(a), the total number of terms at $\alpha = 3 \times 10^{-2}$ is $14$, indicating that {\it Rheo}-SINDy yields a relatively sparse solution from our training dataset. 

Figure~\ref{Fig04} displays the identified coefficient values at the nearly optimal $\alpha$ determined in Fig.~\ref{Fig03}. We found that the lower-order terms are mainly responsible for reproducing the FENE-dumbbell dynamics under extensional flow. Nevertheless, a closer look at the inset in Fig.~\ref{Fig04} makes us realize that the ${\rm tr}_{(1)} {\boldsymbol \tau} = ({\rm tr} {\boldsymbol \tau}) {\boldsymbol \tau}$ terms are systematically obtained (but with small coefficient values). Since Eq.~\eqref{f_FENE_and_tau} shows a clear relation between $f_{\rm FENE}$ and ${\rm tr} {\boldsymbol \tau}$, we can understand that these terms are essential to reproduce the dynamics of the FENE dumbbell model. We also found that, compared to the coefficient values of $\kappa_{yy}$ in the $\overset{\triangledown}{\tau}_{yy}$ equation and $\kappa_{zz}$ in the $\overset{\triangledown}{\tau}_{zz}$ equation, the $\kappa_{xx}$ term in the $\overset{\triangledown}{\tau}_{xx}$ equation has a significant coefficient value. Although Eq.~\eqref{FENE_P_stress_expression_final} shows that the SINDy algorithm naturally identifies this $\kappa_{xx}$ term in the $\overset{\triangledown}{\tau}_{xx}$ equation, its coefficient value is larger than the theoretical value ($=2$) for the FENE-P dumbbell model. The significant coefficient value of the $\kappa_{xx}$ term in the $\overset{\triangledown}{\tau}_{xx}$ equation appears presumably because we set the $x$-direction always to be specified as the extension direction (cf. Eqs.~\eqref{UE}--\eqref{BE}). Using a more symmetric dataset, such as stress data from uniaxial extensional flows in three different extensional directions, would improve the performance of approximate CMs. This finding provides an important insight for developing better models using the {\it Rheo}-SINDy framework.

Figure~\ref{Fig05} shows the performance of the identified approximate CM for oscillatory extensional flows. The comparison between the results obtained by the approximate CM and FENE dumbbell model shows that the approximate CM reasonably reproduces the BD simulation results for the FENE dumbbell model. While we designed the library matrix with the assistance of the analytical solution of the FENE-P dumbbell model, the time evolutions of the stresses obtained by our approximate model are closer to those of the FENE dumbbell model, rather than those of the FENE-P dumbbell model. Figure~\ref{Fig05}(c) shows small deviations in the predictions for the minor stress components (i.e., $\tau_{xx}$ and $\tau_{yy}$). These deviations are due to our training algorithm, which is designed to reduce overall error and is more adaptive to the significant stress component. An accurate description of small-magnitude stress components related to the linear stress response should be the focus of our future work. 

We also show the performance of the approximate CM on steady uniaxial extensional flows not included in the training data. Figure~\ref{Fig06} compares the results obtained by integrating our approximate CM with those from the FENE and FENE-P dumbbell models. The time evolution of stresses predicted by our approximate model more closely follows that of the FENE dumbbell model than that of the FENE-P model, consistent with the trend shown in Fig.~\ref{Fig05}. Comparing the predictions of our model with those of the FENE dumbbell model, while the agreement is only fair ($\sim14\%$ deviation at the steady state) at the lower strain rate ($\dot \epsilon = 4$), it becomes good ($\sim1\%$ deviation at the steady state) at the higher strain rate ($\dot \epsilon = 8$). Furthermore, although we can nicely predict the major stress component, especially for the larger strain rate, the minor stress component shows a deviation highlighted by the insets of Fig.~\ref{Fig06}(a) and (b). The adjustment to the larger extension rate is consistent with the discussion in Fig.~\ref{Fig05}. Figure~\ref{Fig07} compares the results for the larger strain rates than Fig.~\ref{Fig06} obtained by integrating our approximate CM and the FENE dumbbell model. Since we used oscillatory extensional flows with the maximum amplitude of $\epsilon_0 = 8$ for training, these results are extrapolations of the training data with respect to both flow history and strain rate. Even at the maximum strain rate ($\dot \epsilon = 30$), the error at steady state was approximately $8\%$, indicating that our CM can reasonably address the extrapolated situations. Despite the discrepancies observed in Figs.~\ref{Fig05}, \ref{Fig06}, and \ref{Fig07}, our model achieves a substantially lower computational cost than the FENE-dumbbell model. For example, when $\dot \epsilon = 30$, the BD simulation required a tiny time step ($< 10^{-4}$) and a large number of dumbbells. In contrast, our model requires solving only three deterministic differential equations using flexible numerical integration methods.

Our data-driven approach will help efficiently predict transport phenomena and processes governed by extensional flow, such as capillary thinning~\cite{Connell2022,Wang2025} and melt spinning.~\cite{Sato2017,Xu2023,Xu2024} Our current training data includes several stress trajectories under uniaxial, planar, and biaxial flows up to $t = 10 \lambda$, with $\lambda$ being the relaxation time. It is currently challenging to obtain these data solely through experiments (although a promising approach, LAOE, has been proposed by Haward and coworkers~\cite{Recktenwald2025}). However, numerical simulations can address uniaxial, planar, and biaxial flows.~\cite{Murashima2018,Koide2025} Obtaining stress trajectories under oscillatory extensional flows would not be a difficult task for some coarse-grained models. Nevertheless, we should further refine our approach to address data obtained by more complex models than the dumbbell model. For example, our data-driven approach should be robust to noise and address multiple relaxation modes, as in more sophisticated models. We plan to develop our approach in this direction. 
\section{CONCLUSIONS}
This study examined the data-driven approach to derive constitutive models (CMs) under extensional flow. We relied on a simple data-driven approach based on sparse identification of nonlinear dynamics (SINDy). SINDy assumes that a product of library and coefficient matrices expresses the time-series data for the time derivatives of quantities of interest. We utilized our rheological knowledge to design the library matrix. 

Using our data-driven framework, we tested two scenarios: whether it can reproduce a known CM from data generated by that CM, and whether it can derive an approximate CM from data generated by a mesoscopic model whose analytical CM is unknown. We used the Giesekus model for the former, whereas we used the dumbbell model with a finite extensible nonlinear elastic (FENE) spring (i.e., the FENE dumbbell model) for the latter. After confirming that our approach successfully recovered the Giesekus model, we attempted to obtain an approximate CM for the FENE dumbbell model. We used an analytical expression for the FENE-P dumbbell model to design our library matrix, assuming that the FENE dumbbell model is similar to the FENE-P dumbbell model. Our {\it Rheo}-SINDy framework found a relatively simple expression that reproduced the training data generated by the FENE dumbbell model. Test simulations of the identified approximate CM indicate that it can address flow histories and strain rates not included in the training data, demonstrating the potential of {\it Rheo}-SINDy. Based on the baseline results obtained in this study, we plan to refine this method further to make it a valuable tool for rheological modeling.
\section*{ACKNOWLEDGEMENTS}
This work was supported by JST PRESTO Grant Number JPMJPR22O3 (T.S.) and JSPS KAKENHI Grant Number 25K23525 (S.M.).
\appendix
\section{\label{App_A}Constitutive model for the FENE-P model}
This appendix presents the derivation of the constitutive model for the FENE-P dumbbell model.~\cite{Bird1987,Mochimaru1983} Hereafter, we use the unit time as the relaxation time $\lambda$ $(= \zeta/(4h_{\rm eq}))$, the unit length as the equilibrium length of dumbbells $R_{\rm eq}$, and the unit stress as the modulus $G$, and make the time, length, and stress non-dimensionalized by $\tilde t = t / \lambda$, $\tilde {\boldsymbol R} = {\boldsymbol R} / R_{\rm eq}$, and $\tilde {{\boldsymbol \tau}} = {\boldsymbol \tau} / G$, respectively. For simplicity, we will denote non-dimensional variables without a tilde. 

Instead of Eq.~\eqref{FENE}, in the FENE-P dumbbell model, the FENE factor appeared in Eq.~\eqref{spring_strength} is expressed as 
\begin{equation}
  f_{\rm FENE} (t) = \frac{1}{1 - \langle {\boldsymbol R}^2(t) \rangle/R_{\rm max}^2}. 
\label{FENE_P}
\end{equation}
Applying the pre-averaging approximation in Eq.~\eqref{dumbbell_stress}, we can rewrite the stress tensor as 
\begin{align}
{\boldsymbol \tau} (t) 
&= 3 f_{\rm FENE}(t) \langle {\boldsymbol R}(t){\boldsymbol R}(t) \rangle - {\boldsymbol I} \nonumber \\
&= 3 f_{\rm FENE}(t) \boldsymbol C(t) - {\boldsymbol I}, 
\label{fp_dumbbell_stress}
\end{align}
where $\boldsymbol C (t)$ is the conformation tensor, and its time-evolution equation is 
\begin{equation}
\overset{\triangledown}{\boldsymbol C} (t)= - f_{\rm FENE} (t) {\boldsymbol C} (t) + \frac{1}{3} {\boldsymbol I} = -\frac{1}{3} {\boldsymbol \tau} (t). 
\label{C}
\end{equation}
Taking the trace of both sides of Eq.~\eqref{fp_dumbbell_stress} and using the relation $\langle {\boldsymbol R}^2(t) \rangle = R_{\rm max}^2 \{1 - f_{\rm FENE}^{-1}(t)\}$, we can rewrite $f_{\rm FENE} (t)$ as a function of $\boldsymbol \tau (t)$ as
\begin{equation}
f_{\rm FENE} (t) = 1 + \frac{1}{3 R_{\rm max}^2} \left \{ {\rm tr} {\boldsymbol \tau} (t) + 3 \right \}.
\label{f_FENE_and_tau}
\end{equation}
In the following, for simplicity, we omit the notation ``$(t)$'' indicating time dependence. Taking the convected derivative of $\boldsymbol \tau /f_{\rm FENE}$, the time evolution of $\boldsymbol \tau$ can be expressed as
\begin{equation}
\overset{\triangledown}{\boldsymbol \tau} = - f_{\rm FENE}{\boldsymbol \tau} + 2{\boldsymbol D} + \frac{{\rm d} \ln f_{\rm FENE}}{{\rm d}t} ( {\boldsymbol \tau} + {\boldsymbol I} ).
\label{fp_stress_cm}
\end{equation}

From Eq.~\eqref{f_FENE_and_tau}, the time evolution of $\ln f_{\rm FENE}$ can be expressed in terms of ${\rm tr} \boldsymbol \tau$ as
\begin{equation}
\frac{{\rm d}}{{\rm d} t} {\rm tr}  {\boldsymbol \tau} = \left ( 3 R_{\rm max}^2 + {\rm tr}  {\boldsymbol \tau} + 3 \right ) \frac{{\rm d}\ln f_{\rm FENE}}{{\rm d} t}.
\label{f_FENE_and_tr_tau}
\end{equation}
Furthermore, taking trace of Eq.~\eqref{fp_stress_cm} and using Eq.~\eqref{f_FENE_and_tr_tau}, we can obtain
\begin{equation}
\frac{{\rm d}\ln f_{\rm FENE}}{{\rm d} t} 
= \frac{1}{3 R_{\max}^2} \left \{  - f_{\rm FENE} {\rm tr}  {\boldsymbol \tau} + {\rm tr} \left (  {\boldsymbol \tau} \cdot  {\boldsymbol \kappa}^{\rm T} +  {\boldsymbol \kappa} \cdot  {\boldsymbol \tau} \right ) \right \}.
\label{ln_f_FENE}
\end{equation}
We used ${\rm tr} {\boldsymbol D} = 0$ to derive this expression. Combining Eqs.~\eqref{f_FENE_and_tau}, \eqref{fp_stress_cm}, and \eqref{ln_f_FENE} yields
\begin{align}
\overset{\triangledown}{\boldsymbol \tau} &= - \left ( 1 + \frac{1}{R_{\rm max}^2} \right ) {\boldsymbol \tau} + 2{\boldsymbol D} - \frac{1}{3 R_{\rm max}^2} \left ( 1 + \frac{1}{R_{\rm max}^2} \right ) ({\rm tr} {\boldsymbol \tau}) {\boldsymbol I}  \nonumber \\
&\quad - \frac{1}{3R_{\rm max}^2} \left ( 2 + \frac{1}{R_{\rm max}^2} \right ) \left ( {\rm tr} {\boldsymbol \tau} \right ) {\boldsymbol \tau} - \frac{1}{9R_{\rm max}^4} \left ({\rm tr} {\boldsymbol \tau} \right )^2 {\boldsymbol I} \nonumber \\
&\quad + \frac{1}{3 R_{\max}^2} {\rm tr} \left ( {\boldsymbol \tau} \cdot  {\boldsymbol \kappa}^{\rm T} +  {\boldsymbol \kappa} \cdot  {\boldsymbol \tau} \right ) {\boldsymbol I} \nonumber \\ 
&\quad - \frac{1}{9 R_{\rm max}^4} \left ({\rm tr} {\boldsymbol \tau} \right )^2 {\boldsymbol \tau} + \frac{1}{3 R_{\max}^2} {\rm tr} \left (  {\boldsymbol \tau} \cdot  {\boldsymbol \kappa}^{\rm T} +  {\boldsymbol \kappa} \cdot  {\boldsymbol \tau} \right ) {\boldsymbol \tau}, 
\label{FENE_P_stress_expression_final}
\end{align} 
which we used to design our library matrix $\boldsymbol \Theta (\boldsymbol T, \boldsymbol K)$. Note that we can recover the UCM model in the limit of $R_{\rm max} \rightarrow \infty$.  
\section{\label{App_B}Rheo-SINDy test on the FENE-P model}
\begin{figure}[t]
\centering
\includegraphics[width=0.9\columnwidth]{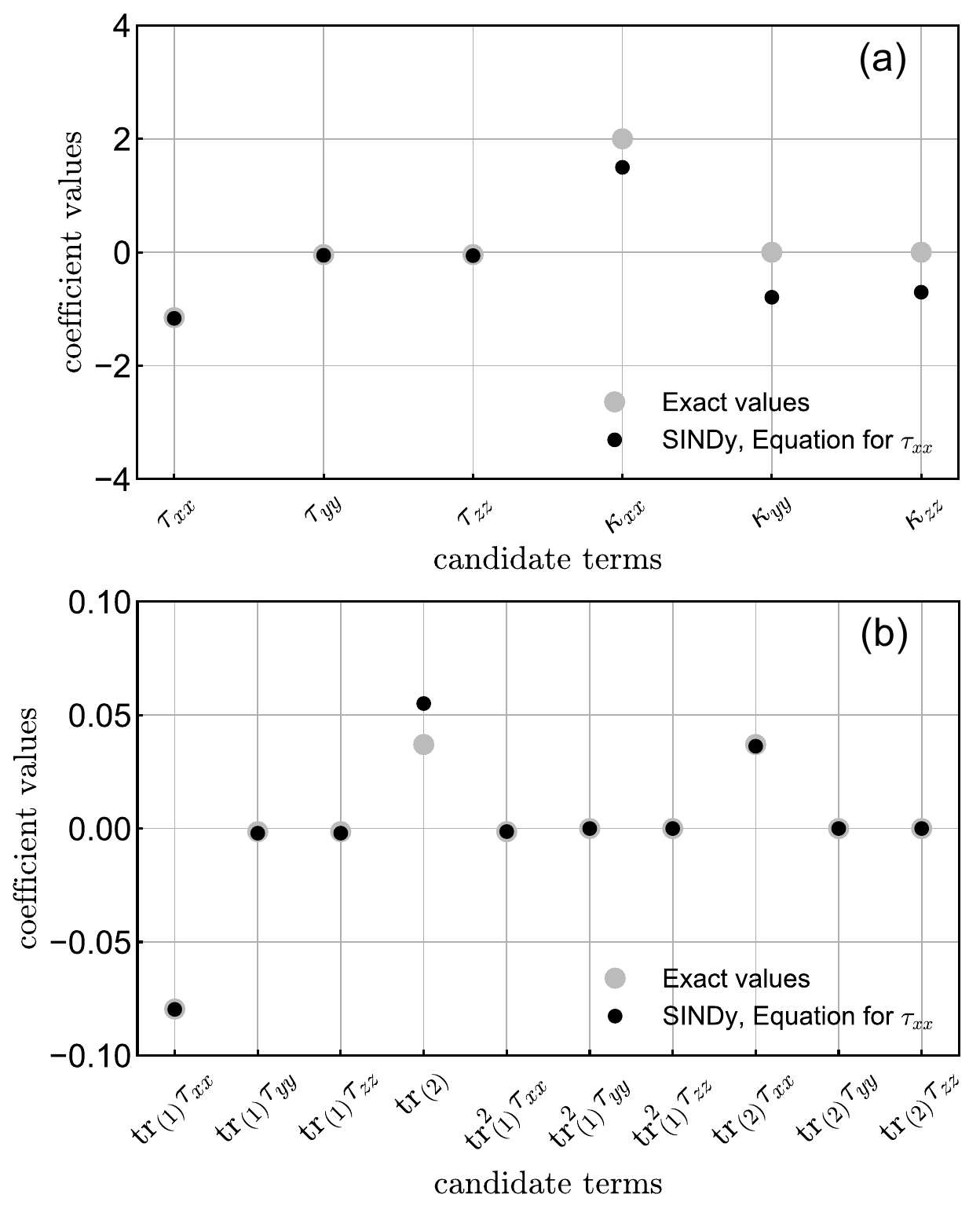}
\caption{The identified coefficient values of the equation for ${\overset{\triangledown}\tau}_{xx}$ for the FENE-P dumbbell model obtained by STRidge with $\alpha = 3 \times 10^{-5}$. The coefficients of the lower and higher order terms are plotted separately in (a) and (b), respectively, on different vertical scales. Black and gray symbols indicate the coefficient values obtained by {\it Rheo}-SINDy and the exact coefficient values, respectively. As in Fig.~\ref{Fig04}, ${\rm tr}_{(1)}$ and ${\rm tr}_{(2)}$ indicate ${\rm tr}_{(1)} = {\rm tr} {\boldsymbol \tau}$ and ${\rm tr}_{(2)} = {\rm tr} ({\boldsymbol \tau} \cdot {\boldsymbol \kappa}^{\rm T} + {\boldsymbol \kappa} \cdot {\boldsymbol \tau})$, respectively.}
\label{Fig08}
\end{figure}
We tested whether {\it Rheo}-SINDy recovers the exact expression of the FENE-P dumbbell model (cf. Eq.~\eqref{FENE_P_stress_expression_final}) under extensional flow. For this purpose, we used STRidge since it was effective in identifying the FENE-P dumbbell model under shear flow in our previous study.~\cite{Sato2025a} Our preliminary test of the FENE-P dumbbell model showed that including training data at high strain rates leads to overfitting to those data and does not necessarily improve the model. Thus, we generated the training data by solving Eqs.~\eqref{FENE_P}--\eqref{C} under the time-dependent $\dot \epsilon (t)$ with $\epsilon_0 \in \{2,4,6\}$ for the uniaxial and planar extensional flows and $\epsilon_0 \in \{1,2,3\}$ for the biaxial extensional flow. Other training conditions, including library design, were the same as those for the FENE dumbbell model. 

Using the same procedure as that for the FENE dumbbell model, we selected the nearly optimal model from a wide range of $\alpha$ values. Figure~\ref{Fig08} compares the coefficient values obtained by {\it Rheo}-SINDy with the exact coefficient values (i.e., Eq~\eqref{FENE_P_stress_expression_final} with $R_{\rm max} = 3$) for the equation for ${\overset{\triangledown}\tau}_{xx}$. Although there are slight differences between the obtained and exact values, mainly in the terms related to the velocity gradient tensor, {\it Rheo}-SINDy can reasonably identify the coefficient values of the other terms. The partial success shown in Fig.~\ref{Fig08} indicates that the {\it Rheo}-SINDy framework can reasonably work for the constitutive model with highly nonlinear terms. 

\end{document}